%% file: Diego_proceedings.tex
\documentclass[slac_one]{revtex4}
\usepackage{graphicx}
\usepackage{fancyhdr}
\usepackage[abs]{overpic}

\usepackage{graphicx}
\usepackage{lscape}
\usepackage{mathrsfs}
\usepackage[abs]{overpic}
\usepackage{epsfig}
\usepackage{rotating}
\usepackage{dcolumn}
\usepackage{color}
\usepackage{eso-pic}
\usepackage{textcomp}
\usepackage{amsmath}
\usepackage{amssymb}
\usepackage{xspace}
\usepackage{relsize}
\usepackage{pstricks}
\usepackage{pst-node}
\usepackage{floatflt}

\input{macro}

\newcommand{\betas}{\ensuremath{\beta_{s}^{J/\psi\phi}}}   	
\newcommand{\betasSM}{\ensuremath{\beta_{s}^{\mathrm{SM}}}}	
\newcommand{\betasNP}{\ensuremath{\beta_{s}^{\mathrm{NP}}}}	

\newcommand{\phis}{\ensuremath{\phi_{s}^{J/\psi\phi}}}   	
\newcommand{\phisSM}{\ensuremath{\phi_{s}^{\mathrm{SM}}}}

\newcommand{\DGSM}{\ensuremath{\Delta\Gamma^{\mathrm{SM}}}}

\begin{document}

\title{Search for New Physics in the \boldmath $B^0_s$ \unboldmath mixing phase at CDF} 

\author{D.~Tonelli, (for the CDF collaboration)}
\affiliation{Fermilab, Batavia, P.O. Box 500, IL 60510-5011, USA}

\begin{abstract}
The Collider Detector at Fermilab (CDF) experiment performed the first measurement of the time-evolution of flavor-tagged \bsjpsiphi\ decays, which probes mixing-induced \CP-violation in the \bs\ sector. Any sizable deviation from zero of the phase \betas, accessible through interference of the $\bar{b}\to \bar{c}c\bar{s}$ quark-level process accompanied or not by $\bs-\abs$ mixing, would be unambiguous indication of physics beyond the Standard Model. I report CDF results obtained in 1.35 \lumifb, a recent extension to a larger dataset corresponding to 2.8 \lumifb, and future projections.
\end{abstract}

\maketitle

\thispagestyle{fancy}


\section{INTRODUCTION} 
Many precise results from several years of successful $B$--factories' running disfavor significant ($\gessim 10\%$) contributions from ``New Physics'' (NP) in tree-dominated \bgmeson\ decays. Agreement with the Standard Model (SM) is also found in higher-order processes such as $K^0$--$\overline{K}^0$ or \bd--\abd\ transitions due to second-order weak interactions (mixing) that involve virtual massive particles and may receive contributions from NP. A less clear picture is available for the \bs\ system. The strength of NP contributions in \bs--\abs\ mixing is constrained by the precise measurement of the oscillation frequency \cita{mixing}, which disfavors large magnitudes of NP amplitudes. However, knowledge of only the frequency  leaves the phase of the mixing amplitude unconstrained. Indeed, possible large NP phases are currently not excluded. The mixing phase is accessible through the time-evolution of \bsjpsiphi\ decays, which is sensitive to the relative phase between the mixing and the $\bar{b}\to \bar{c}c\bar{s}$ quark-level transition, $\betas = \betasSM+\betasNP$. Such phase is responsible for \CP-violation and is $\betasSM=\arg(-V_{ts}V_{tb}^{*}/V_{cs}V_{cb}^{*}) \approx 0.02$ in the SM \cita{globalfit}; any sizable deviation from this value would be unambiguous evidence of NP \cita{theory}. If NP contributes a phase ($\betasNP$), this would also enter $\phis = \phisSM - 2\betasNP$, which is the phase difference between mixing and decay into final states common to \bs\ and \abs, and is tiny in the SM: $\phisSM = \arg(-M_{12}/\Gamma_{12}) \approx 0.004$ \cita{constraint}. The phase \phis\ enters the decay-width difference between light and heavy states, $\Delta\Gamma=\Gamma_L-\Gamma_H=2|\Gamma_{12}|\cos(\phis)$, which is $\DGSM \approx 2|\Gamma_{12}| = 0.096 \pm 0.036$ ps$^{-1}$ in the SM \cita{constraint} and plays a r\^{o}le in \bsjpsiphi\ decays. Since the SM values for \betas\ and \phis\ cannot be resolved with the resolution of current experiments, the following approximation is used: $\phis \approx -2\betasNP \approx -2\betas$, which holds in case of sizable NP contributions. \par This measurement of \betas\ is analogous to the determination of the phase $\beta=\arg(-V_{cd}V_{cb}^{*}/V_{td}V_{tb}^{*})$ in $\bd \to \jpsi\Ks$ decays, except for a few additional complications: the oscillation frequency is about 35 times higher in \bs\ than in \bd\ mesons, requiring excellent decay-time resolution; the decay of a pseudoscalar meson (\bs) into two vector mesons ($\jpsi$ and $\phi$) produces two \CP-even states (orbital angular momentum $L = 0,2$), and one \CP-odd state ($L=1$), which should be separated for maximum sensitivity; and the value of the SM expectation for $\betas$ is approximately $30$ times smaller \cita{bigi-sanda} than the known $\beta$ value \cita{hfag}.
\section{SIGNAL SELECTION}
\begin{figure}
\begin{overpic}[bb= 0 0 567 545, scale=0.39, clip=true]{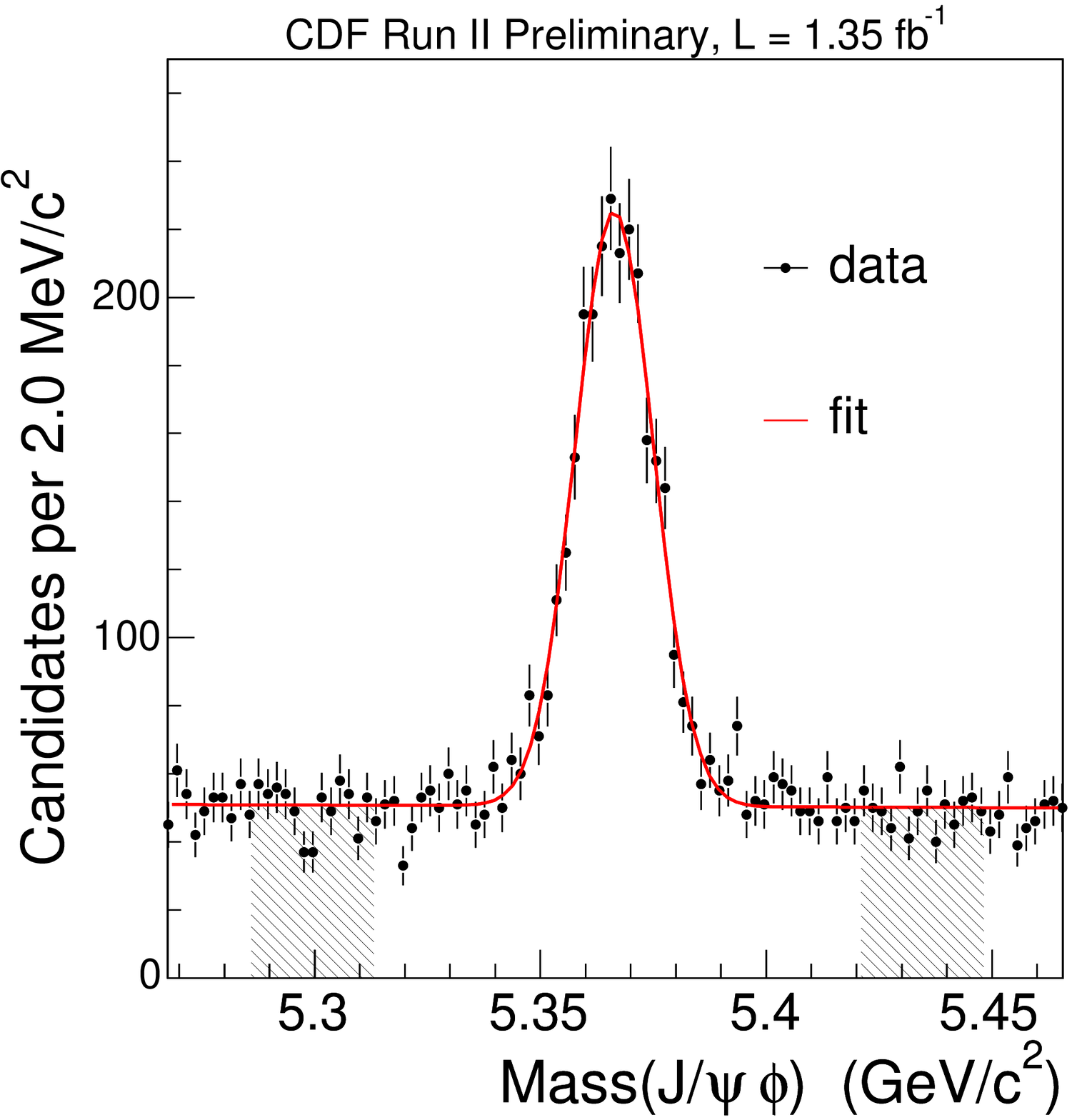}
\put(40,200){(a)}
\end{overpic}
\begin{overpic}[bb= 0 6 567 500, scale=0.432, clip=true]{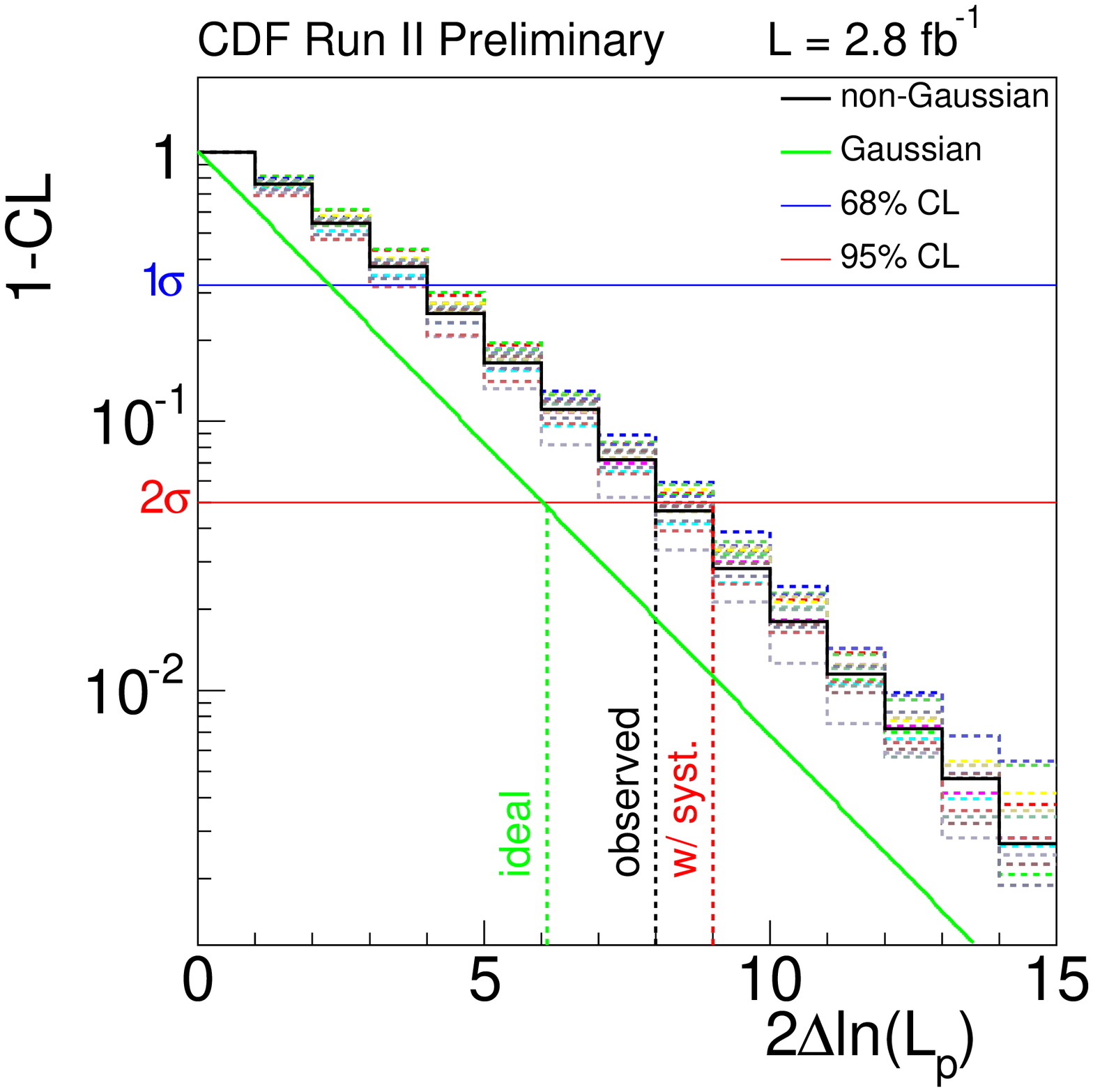}
\put(55,200){(b)}
\put(160, 7){\Large $-$}
\end{overpic}
\caption{\label{fig:yield} Mass distribution with fit projection and sideband regions overlaid (a). Distribution of $-2\Delta\ln(L_p)$ observed in pseudo-experiments (solid black line) compared with the nominal $\chi^2$ distribution (solid green line) (b). The effect of sampling the nuisance parameters within $5\sigma$ of their estimates in data is shown by the dashed colored lines. The vertical lines indicate the different values of $-2\Delta\ln(L_p)$ to be used in the ideal $\chi^2$ case (green), in the observed distribution (black), and in the observed distribution including systematic uncertainties (red), to obtain a 95\% C.L. region. The distribution shown corresponds to a specific $(\betas, \Delta\Gamma)$ point, but similar distributions have been observed across the whole space.}
\end{figure}

The CDF experiment at the Fermilab Tevatron performed the first measurement of the time-evolution of flavor-tagged $\bs \to \jpsi(\to\mu^+\mu^-) \phi(\to K^+K^-)$ decays \cita{PRL_betas_CDF}. These were reconstructed in \pap\ collision data corresponding to a time-integrated luminosity of 1.35 \lumifb. 
Events enriched in \jpsi\ decays are selected by a trigger that requires the spatial matching between a pair of two-dimensional, oppositely-curved, tracks 
in the multi-wire drift chamber (coverage $|\eta|<1$) and their extrapolation outward to track-segments reconstructed in the muon detectors
(drift chambers and scintillating fibers). In the offline analysis, a kinematic fit to a common space-point is applied between the candidate \jpsi\ and another pair of tracks consistent with being kaons originated from a $\phi$ meson decay. The measurement of specific energy loss by ionization in the drift chamber (\dedx) provides $1.5\sigma$ separation between charged kaons and pions with momenta $p>2~\pgev$. At lower momenta, scintillators bars surrounding the chamber measure arrival times of charged particles (time-of-flight, TOF) with approximately 110 ps resolution, providing separation between kaons and pions in excess of $2\sigma$. An artificial neural network trained on simulated data (to identify signal, $S$) and \bs\ mass sidebands (for background, $B$) is used for an unbiased optimization of the selection. The quantity $S/\sqrt{S+B}$ is maximized using kinematic and particle identification (PID) information. Attempts of using the average statistical resolution on $\betas$ observed in ensembles of pseudoexperiments as figure of merit were inconclusive because of irregularities of the likelihood (see below). Discriminating observables include kaon-likelihood, from the combination of \dedx\ and TOF information; transverse momenta of the \bs\ and $\phi$ mesons; the $K^+K^-$ mass; and the quality of the vertex fit. The final sample contains approximately 2000 signal events over a comparable background (\fig{yield} (a)). Seven layers of silicon sensors extending radially up to 22 cm, and the drift chamber that provides 96 measurements between 30 and 140 cm, all immersed in the 1.4 T axial magnetic field, provide a mass resolution of approximately 10 \massmev\ on the \bsjpsiphi\ peak. 
\section{FITTING THE TIME EVOLUTION}
The sensitivity to the mixing phase is enhanced if the evolution of \CP-even eigenstates, \CP-odd eigenstates, and their interference is separated. CDF uses the angular distributions of final state particles to statistically determine the \CP-composition of the signal. The angular distributions are studied in the transversity basis, which allows a convenient separation between \CP-odd and \CP-even terms in the equations of the time-evolution \cita{transversity}.\par Sensitivity to the phase increases if the evolutions of bottom-strange mesons produced as \bs\ or \abs\ are studied independently.  The time development of flavor-tagged decays contains terms proportional to $\sin(2\betas)$, reducing the ambiguity with respect to the untagged case ($\propto |\sin(2\betas)|$). Building on techniques used in the \bs\ mixing frequency measurement \cita{mixing}, the production flavor is inferred using two classes of algorithms. Opposite-side tags exploit \bab\ pair production, the dominant source of \bhadrons\ at the Tevatron, and estimate the production flavor from the charge of decay products ($e$, $\mu$, or jet) of the $b$--hadron produced from the other $b$--quark in the event. Same-side tags rely on the charges of associated particles produced in the fragmentation of the $b$--quark that hadronizes into the candidate \bs\ meson. The tagging power, $\epsilon D^2 \approx 4.5\%$,  is the product of an efficiency $\epsilon$, the fraction of candidates with a flavor tag, and the square of the dilution $D=1-2w$, where $w$ is the mistag probability. Multiple tags, if any, are combined as independent. The proper time of the decay and its resolution are known on a per-candidate basis from the position of the decay vertex, which is determined with an average resolution of approximately 27 \mum\ (90 fs$^{-1}$) in \bsjpsiphi\ decays, owing to the first layer of the silicon detector at 1.6 cm radius from the beam.\par
Information on \bs\ candidate mass and its uncertainty, angles between final state particles' trajectories (to extract the \CP-composition), production flavor, and decay length and its resolution are used as observables in a multivariate unbinned likelihood fit of the time evolution that accounts for direct decay amplitude, mixing followed by 
the decay, and their interference. Direct \CP-violation is expected to be small and is not considered. The fit determines the phase $\betas$, the decay-width difference $\Delta\Gamma$, and 25 other ``nuisance'' parameters ($\vec{\nu}$). These include the mean \bs\ decay-width ($\Gamma = (\Gamma_L + \Gamma_H)/2$), the squared magnitudes of linear polarization amplitudes ($|A_0|^2$, $|A_{\parallel}|^2$, $|A_{\perp}^2|$), the \CP-conserving (``strong'') phases ($\delta_{\parallel} = \arg(A_{\parallel} A_{0}^{*})$, $\delta_{\perp} = \arg(A_{\perp}A_{0}^{*})$), and others.
The acceptance of the detector is calculated from a Monte Carlo simulation and found to be consistent with observed angular distributions of random combinations of four tracks in data; the angular-mass-lifetime model was validated by measuring lifetime and polarization amplitudes in 7800 \bdjpsikstar\ decays, which show angular features similar to the \bs\ sample: $c\tau(\bd) = 456 \pm 6 \stat \pm 6 \syst \mum$, $|A_0|^2 = 0.569 \pm 0.009 \stat \pm 0.009 \syst$, $|A_\parallel|^2 = 0.211 \pm 0.012 \stat \pm 0.006 \syst$, $\delta_\parallel = -2.96 \pm 0.08 \stat \pm 0.03 \syst$, and $\delta_\perp = 2.97 \pm 0.06 \stat \pm 0.01 \syst$. The results, consistent and competitive with most recent $B$--factories' results \cita{bdjpsikstar}, support the reliability of the model. Additional confidence is provided by the precise measurement of lifetime and width-difference in untagged \bsjpsiphi\ decays \cita{untagged}.  
\section{STATISTICAL ISSUES}
Tests of the fit on simulated samples show biased, non-Gaussian distributions of estimates and multiple maxima, because the likelihood is invariant under the transformation $\mathcal{T} = (2\betas\to\pi-2\betas, \Delta\Gamma\to-\Delta\Gamma, \delta_\parallel\to2\pi-\delta_\parallel,\delta_\perp\to\pi-\delta_\perp)$, and the resolution on \betas\ was found to depend crucially on the true values of \betas\ and $\Delta\Gamma$.
CDF quotes therefore a frequentist confidence region in the ($\betas, \Delta\Gamma$) plane rather than point-estimates for these parameters. Obtaining a correct and meaningful region from a multidimensional likelihood is challenging: one should construct the full 27-dimensional region, a difficult task computationally, and project it onto the ($\betas,\Delta\Gamma$) plane. The choice of the ordering algorithm is critical to prevent the projection from covering most of the ($\betas,\Delta\Gamma$) space, yielding a scarcely informative result. A common approximate method is to replace the likelihood, $L(\betas,\Delta\Gamma, \vec{\nu})$, with the \emph{profile} likelihood, $L_{p}(\betas,\Delta\Gamma, \hat{\vec{\nu}})$. For every point in the ($\betas,\Delta\Gamma$) plane, $\hat{\vec{\nu}}$ are the values of nuisance parameters that maximize the likelihood. Then $-2\Delta\ln(L_p)$ is typically used as a $\chi^2$ variable to derive confidence regions in the two-dimensional space ($\betas,\Delta\Gamma$). However, the simulation shows that in the present case the approximation fails: the resulting regions contain the true values with lower probability than the nominal confidence level (C.L.) because the  $-2\Delta\ln(L_p)$ distribution has longer tails than a $\chi^2$, and is not even independent of the true values of the nuisance parameters (\fig{yield} (b)). A full confidence region construction is therefore needed, using simulation of a large number of pseudo-experiments to derive the actual distribution of $-2\Delta\ln(L_p)$, with a potential for an excessive weakening of the results from systematic uncertainties. However, in a full confidence limit construction, the use of $-2\Delta\ln(L_p)$ as ordering function is close to optimal for limiting the impact of systematic uncertainties \cita{stat}. With this method, CDF is able to rigorously account for the effect of systematic uncertainties just by randomly sampling a limited number of points in the space of all nuisance parameters: a specific value $(\betas, \Delta\Gamma)$ is excluded only if it can be excluded for any assumed value of the nuisance parameters within $5\sigma$ of their estimate on data. The result is a \betas--$\Delta\Gamma$ contour that is the truly two-dimensional projection of the full, 27-dimensional confidence region.
\begin{figure}
\begin{overpic}[width=0.475\textwidth,angle=0]{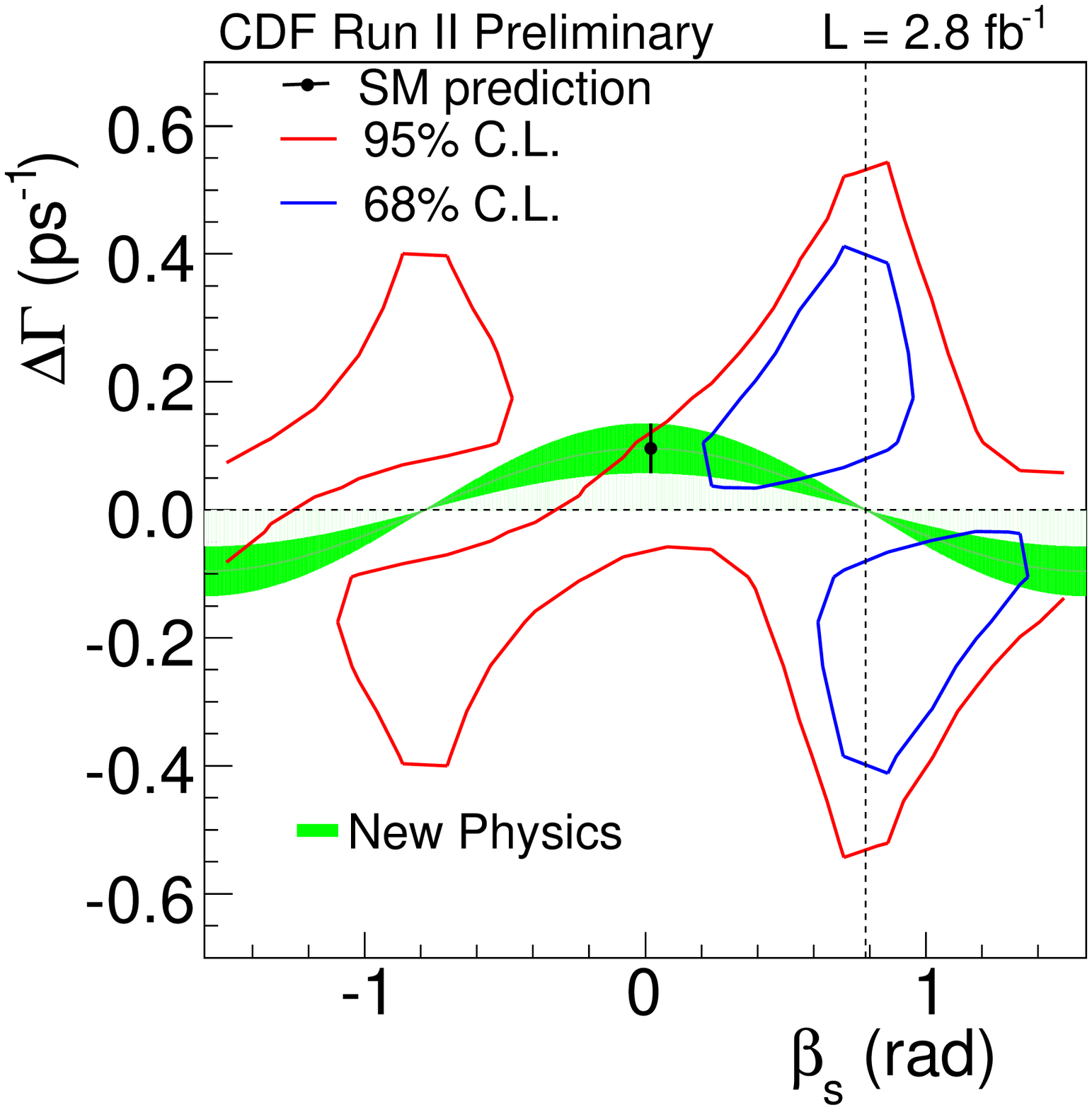}
\put(200,200){(a)}
\end{overpic}
\begin{overpic}[bb= 75 -13.5 437 312, scale=0.72, clip=true]{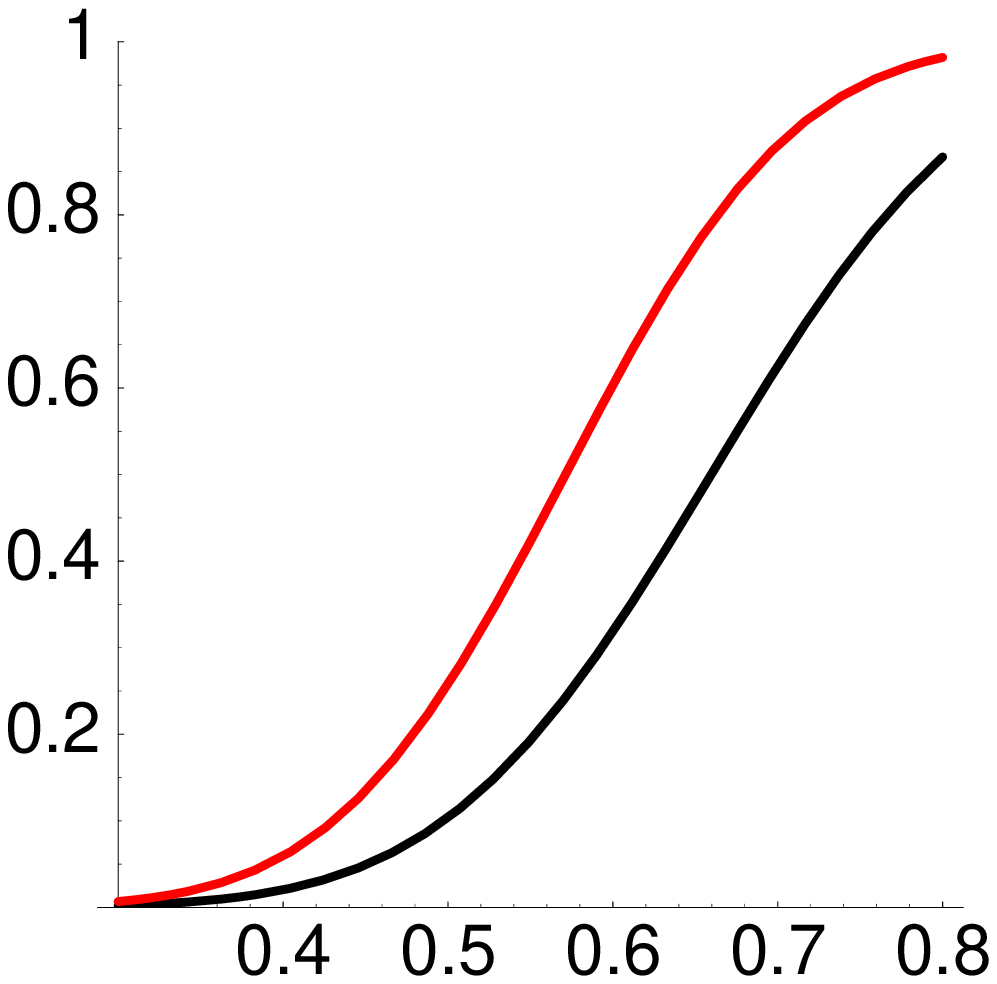}
\put(40,200){(b)}
\put(200,3){\large \boldmath $\betas$\unboldmath}
\end{overpic}
\caption{\label{fig:contours}Confidence region in the $(\betas,\Delta\Gamma)$ plane obtained with 2.8 \lumifb\ of CDF data (a). The green band is the region allowed by any NP contribution not entering $|\Gamma_{12}|$, and assuming $2|\Gamma_{12}|= 0.096 \pm 0.036$ ps$^{-1}$ \cita{constraint}. Fraction of CDF-equivalent experiments that would observe a $5\sigma$ deviation from the SM as a function of the value of \betas\ in a sample corresponding to 6 \lumifb\ (black line) or 8 \lumifb\ (red line) of integrated luminosity (b).}
\end{figure}
\section{RESULTS}
The results on 1.35 \lumifb\ show a 1.5$\sigma$ fluctuation with respect to the SM values \cita{PRL_betas_CDF}. Considering $\Delta\Gamma$ as an additional nuisance parameter, the 68\% C.L. allowed region for the mixing phase is $0.16 < \betas < 1.41$, which restricts to $\betas \in[0.12,0.68]\cup[0.89,1.45]$ assuming no NP contributions in $|\Gamma_{12}|$ (\ie\ constraining $2|\Gamma_{12}|$ to $0.096\pm0.036$ ps$^{-1}$ \cita{constraint}). This result has been confirmed by the D\O\ Collaboration \cita{betas_D0}, which observed a consistent fluctuation in an analysis where the two-fold symmetry of the likelihood is removed by assuming an additional theoretical constraint between strong phases of \bsjpsiphi\ and \bdjpsikstar\ decays \cita{gronau}. After removing this assumption, CDF and D\O\ results can be combined yielding a $2.2\sigma$ fluctuation with respect to the SM  and the following 68\% C.L. range: $\betas \in[0.24,0.57]\cup[0.99,1.33]$ \cita{hfag}. 
\par CDF has reported at this conference a partial extension of the analysis to a larger sample, corresponding to 2.8 \lumifb. This is approximately equivalent to 2.0 \lumifb\ effective luminosity,  because the calibration of \dedx\ and TOF was unavailable for the whole sample and PID information is not used in the selection, nor in flavor tagging for the second half of the dataset. More than 3200 decays are reconstructed, but approximately 4000 are expected when PID will be available in the selection. \Fig{contours} (a) shows the results. The two regions symmetric with respect to the $(\pi/4, 0)$ point reflect the symmetry of the likelihood, which cannot determine from data if $\cos(\delta_\perp)<0$ and $\cos(\delta_\perp -\delta_\parallel)>0$ (corresponding to the $\Delta\Gamma>0$ solution) or \emph{viceversa} ($\Delta\Gamma<0$). The fluctuation with respect to the SM is confirmed and strengthened, reaching the 1.8$\sigma$ level. The updated analysis restricts the allowed regions for the phase to the range $0.28 < \betas < 1.29$ at the 68\% C.L.\par Although the observed deviations are not yet significant, the pattern of independent results showing consistent fluctuations in the same direction is promising in view of the analysis of the full dataset, expected to reach approximately 6 \lumifb\ by year 2009, or 8 \lumifb\ by 2010, if Run II of the Tevatron will be extended. \Fig{contours} (b) shows the probability of a $5\sigma$ exclusion of the SM at CDF as a function of the value of \betas\ in these two scenarios and assuming $\Delta\Gamma = 0.1$ ps$^{-1}$. This extrapolation, which assumes no external constraints and no improvements in the analysis, is conservative: CDF is improving the analysis, with significantly increased tagging power, a 50\% additional signal collected by other triggers, and the possibility to resolve the strong-phases ambiguity using data \cita{strong}; tight external constraints (\eg\ on the \bs\ lifetime) can be applied, and CDF and D\O\ results will be combined for maximum Tevatron sensitivity. As happened in the past, deviations from expectations in measurements of lower-energy processes may indicate NP prior to direct discovery of new resonances, as those expected in the forthcoming run of the Large Hadron Collider \cita{NP}.
\begin{acknowledgments}
I would like to thank my CDF colleagues, who provided valuable suggestions during the preparation of this manuscript, and M.~Gronau for spotting an error in the definition of the strong phases.
\end{acknowledgments}


\end{document}

%% file: macro.tex
\def \rightdownarrow
 {\kern.3em
 \rule[.5ex]{.15mm}{2ex}
 {\mbox{$\kern-0.1em{\longrightarrow}$}}      
 }

\def\lessim{\mathrel {\vcenter {\baselineskip 0pt \kern 0pt  
\hbox{$<$} \kern 0pt \hbox{$\sim$} }}}

\def\gessim{\mathrel {\vcenter {\baselineskip 0pt \kern 0pt   
\hbox{$>$} \kern 0pt \hbox{$\sim$} }}}

\newcommand{\ie}{i.~e.,}					
\newcommand{\eg}{e.~g.,}					

\newcommand{\lumifb}{\mbox{fb$^{-1}$}}				

\newcommand{\mum}{\mbox{$\mu$m}}				

\newcommand{\tev}{\ensuremath{\mathrm{Te\kern -0.1em V}}}
\newcommand{\gev}{\ensuremath{\mathrm{Ge\kern -0.1em V}}}	
\newcommand{\mev}{\ensuremath{\mathrm{Me\kern -0.1em V}}}	
\newcommand{\kev}{\ensuremath{\mathrm{ke\kern -0.1em V}}}	
\newcommand{\massmev}{\mbox{\mev/$c^2$}}			
\newcommand{\pgev}{\mbox{\gev/$c$}}				

\newcommand{\stat}{\ensuremath{\mathit{~(stat.)}}}		
\newcommand{\syst}{\ensuremath{\mathit{~(syst.)}}}		

\newcommand{\CP}{\ensuremath{\mathsf{CP}}}			

\newcommand{\pap}{\proton\antiproton}			

\newcommand{\proton}{\ensuremath{\mathrm{p}}}
\newcommand{\antiproton}{\ensuremath{\bar{\rm{p}}}}

\newcommand{\bab}{\ensuremath{b\bar{b}}}

\newcommand{\Ks}{\ensuremath{K^0_S}}

\newcommand{\bd}{\ensuremath{B^{0}}}				
\newcommand{\bs}{\ensuremath{B^{0}_s}}				

\newcommand{\abd}{\ensuremath{\overline{B}^{0}}}		
\newcommand{\abs}{\ensuremath{\overline{B}^{0}_s}}		

\newcommand{\bhadrons}{\mbox{$b$--hadrons}}			

\newcommand{\bgmeson}{\mbox{$b$--meson}}				

\newcommand{\bdjpsikstar}{\ensuremath{\bd \to  \jpsi K^{*0}}}

\newcommand{\bsjpsiphi}{\ensuremath{\bs \to  \jpsi \phi}}

\newcommand{\jpsi}{\ensuremath{J/\psi}}

\newcommand{\Fig}[1]{Figure \ref{fig:#1}}

\newcommand{\fig}[1]{fig.~\ref{fig:#1}}

\newcommand{\cita}[1]{\cite{#1}}

\newcommand{\dedx}{\ensuremath{\mathit{dE/dx}}}

\newcommand{\etal}{\textit{et al.}}

\def\babar{\mbox{\slshape B\kern-0.1em{\smaller A}\kern-0.1em B\kern-0.1em{\smaller A\kern-0.2em R}}}